\documentclass[sn-nature]{sn-jnl}

\usepackage{graphicx}%
\usepackage{multirow}%
\usepackage{amsmath,amssymb,amsfonts}%
\usepackage{amsthm}%
\usepackage{mathrsfs}%
\usepackage[title]{appendix}%
\usepackage{xcolor}%
\usepackage{textcomp}%
\usepackage{manyfoot}%
\usepackage{booktabs}%
\usepackage{algorithm}%
\usepackage{algorithmicx}%
\usepackage{algpseudocode}%
\usepackage{listings}%
\usepackage[version=4]{mhchem}
\usepackage{comment}
\usepackage{geometry}

\begin{document}

\title[Article Title]{Nanoscale Mapping of Magnetic Auto-oscillations with a single Spin Sensor}


\author*[1,2]{\fnm{Toni} \sur{Hache}}\email{t.hache@fkf.mpg.de}

\author[1]{\fnm{Anshu} \sur{Anshu}}

\author[2]{\fnm{Tetyana} \sur{Shalomayeva}}



\author[2]{\fnm{Rainer} \sur{Stöhr}}

\author[1,3]{\fnm{Klaus} \sur{Kern}}

\author[1,2,4]{\fnm{Jörg} \sur{Wrachtrup}}

\author*[1,4]{\fnm{Aparajita} \sur{Singha}}\email{a.singha@fkf.mpg.de}

\affil*[1]{\orgname{Max Planck Institute for Solid State Research}, \orgaddress{\street{Heisenbergstr. 1}, \city{Stuttgart}, \postcode{70569}, \country{Germany}}}

\affil[2]{\orgdiv{3rd Institute of Physics and Research Center SCoPE}, \orgname{University of Stuttgart},  \orgaddress{\city{Stuttgart}, \postcode{70049}, \country{Germany}}}


\affil[3]{\orgdiv{Institute de Physique}, \orgname{École Polytechnique Fédérale de Lausanne}, \orgaddress{\city{Lausanne}, \postcode{CH-1015}, \country{Switzerland}}}

\affil[4]{\orgname{Center for Integrated Quantum Science and Technology IQST, University of Stuttgart, \city{Stuttgart}, \postcode{70049}, \country{Germany}}}


\abstract{Magnetic auto-oscillations are damping-compensated magnetization precessions. They can be generated in spin Hall nano-oscillators (SHNO) among others. Current research on these devices is dedicated to create next generation energy-efficient hardware for communication technologies. However, the underlying physics governing the formation of auto-oscillation modes, their output power and line width in a single SHNO device have remained elusive so far. We image the sources of magnetic auto-oscillations in a metallic SHNO using a single spin quantum sensor. We directly measure the microwave field generated by an auto-oscillation spot at the nanoscale by driving the electron spin resonance transition of the sensor spin, enabling faster acquisition speed (100 ms/pixel). 
Instead of being defined by the points of the largest antidamping only, we experimentally demonstrate for the first time with quantitative magnetometry that the auto-oscillation spots are determined by the positions of the magnetic field minima. The latter act as local potential wells for confining spin-waves, thus supporting large amplitude auto-oscillations. By comparing the magnitude of the magnetic stray field at these spots, we decipher the different frequencies of the auto-oscillation modes. 
The insights gained regarding the interaction between auto-oscillation modes and spin-wave potential wells enable advanced engineering of real devices.}

\keywords{PL map, auto-oscillation, spin Hall effects, nitrogen-vancancy center, nano-oscillator, nonlinear oscillator}



\maketitle

The efficient excitation, manipulation and readout of spin waves is one of the main foci of current research for creating highly energy-efficient hardware for communication technologies\cite{Finocchio_2024,Dieny:2020aa, Kimel_2022, Kiselev:2003aa, Demidov:2012aa, Chumak:2015aa,PhysRevApplied.17.014003,Demidov:2020aa,Flebus:2024aa}. Large spin-wave amplitudes are achieved in various nano-oscillator devices\cite{PhysRevB.102.054422,https://doi.org/10.48550/arxiv.2301.09228,doi:10.1063/5.0008988,PhysRevLett.123.057204,Haidar:2019aa,Divinskiy:2018aa,C6NR07903B,Divinskiy:2017aa,Spin-Torque_and_Spin-Hall_Nano-oscillators,doi:10.1063/1.4901027,PhysRevLett.109.186602,Yang:2015aa,Nanowire_spin_torque_oscillator_driven_by_spin_orbit_torques,Demidov:2015aa} via damping compensation resulting in magnetic auto-oscillations. Despite their different geometries, the fundamental principle remains very similar. The interaction of a spin current with a ferromagnetic material generates a spin-transfer or spin-orbit torque compensating the damping\cite{ISI:000264630400001, SLONCZEWSKI1996L1}.

As the auto-oscillation frequencies typically lie within GHz range, such nano-oscillators are regarded as miniaturized frequency converters with various mechanisms to control the frequency like the dc current, external magnetic field or synchronization to external stimuli\cite{PhysRevApplied.13.054009,doi:10.1063/1.5082692,PhysRevApplied.19.034070,Synchronization_of_spin_Hall_nano-oscillators_to_external_microwave_signals,Wagner:2018aa,Injection_Locking_and_Phase_Control_of_Spin_Transfer_Nano-oscillators}. These aspects render them as attractive candidates for nanoscale microwave voltage and spin wave sources, neuromorphic computing hardware\cite{Christensen_2022,Grollier:2020aa} or localized microwave field generators, for manipulating single spins in quantum technologies. 
Recently, it was shown that mutual synchronization of several devices results in higher output power and reduced linewidth\cite{Zahedinejad:2020aa, Awad:2017aa, Kumar:2023aa}. However, understanding of the underlying physics of a single device is still largely missing, which is crucial for further improvement of the oscillator properties. In particular, the spatial distribution of the auto-oscillation modes has so far been studied only via micromagnetic simulations\cite{PhysRevApplied.9.014017,Spin-Hall-nano-oscillator:-A-micromagnetic-study,Micromagnetic_study_of_auto-oscillation_modes_in_spin-Hall_nano-oscillators,Mutual_synchronization_of_nanoconstriction-based_spin_Hall_nano-oscillators_through_evanescent_and_propagating_spin_waves}. Furthermore, the origin and mode control of the auto-oscillations in single devices are yet to be achieved. It is expected that such optimized mutually-synchronized oscillators would result in an even higher output power and coherence. 

Due to the small dimension of the nano-oscillators, widely used optical techniques can’t reveal the auto-oscillation mode profile due to size of the diffraction-limited laser spot. Moreover, stroboscopic high-resolution x-ray-techniques can’t be applied due to the missing phase relation between applied dc current and the auto-oscillations.

So far, these difficulties have hindered the experimental observation of auto-oscillation origins in nano-oscillators. Here, we circumvent the above mentioned issues by detection of the microwave field generated by the auto-oscillations with a single spin scanning-probe consisting of a nitrogen-vacancy (NV) center with a spatial resolution of $<$100 nm. Our approach allows us fast, all-optical mapping of distinct auto-oscillation modes. Combing these with additional NV magnetometry measurements\cite{Rondin_2014, doi:10.1063/5.0084910,10.1063/1.2943282,Balasubramanian:2008aa,Maletinsky:2012aa} and simulations we unravel the origin behind the characteristic frequencies of auto-oscillations.



\subsection*{SHNO and Measurement geometry}

\begin{figure}[t]
\begin{center}
\scalebox{1}{\includegraphics[width=18cm, clip]{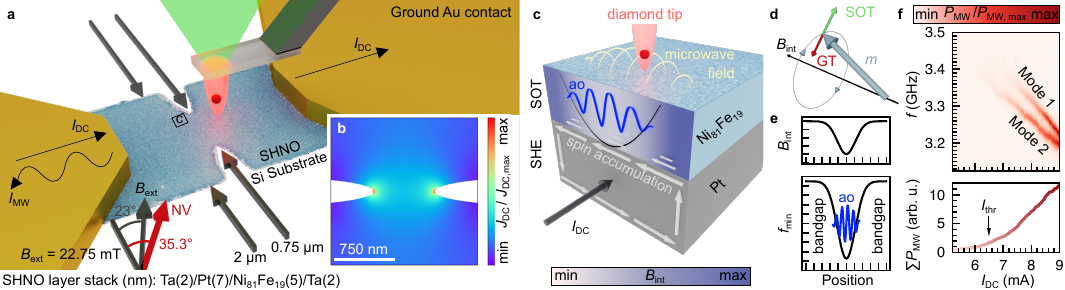}}
\caption{\label{fig1} Measurement overview and electrical characterization of the SHNO device. (a) Sample geometry and schematic representation of the electrical and optical measurements. (b) The simulated dc current density $J_{\textnormal{DC}}$ distribution reveals maximum values at the edges of the constriction. 
(c) A pure spin current generated via SHE in Pt generates a SOT in \ce{Ni81F19}. This generates magnetic auto-oscillations which generate a microwave field interacting with the single spin sensor. (d) The SOT compensates the GT resulting in auto-oscillations. (e) Local magnetic field minima create spin wave potential wells which confine the auto-oscillations.
(f) Top: Auto-oscillation spectra electrically measured as a function of dc currents exhibit the presence of two auto-oscillation modes with characteristic microwave frequencies. Bottom: Integrated auto-oscillation power}
\end{center}
\end{figure}

We utilize a spin Hall nano-oscillator (SHNO) in order to generate magnetic auto-oscillations. Fig. 1(a) shows the schematic of the SHNO device (blue) with a well-defined constriction of $d=750$ nm at the center which forms spots with high current densities as shown in Fig. 1(b). 
The spin Hall effect (SHE)\cite{DyakonovPerrel,ISI:000082242600034,ISI:000324930200001} within the Pt layer (Fig. 1(c)) generates a pure spin current perpendicular to the applied charge current, resulting in the injection of electrons with well defined spin polarization into the adjacent soft ferromagnetic \ce{Ni81F19}  layer. The dc current polarity controls the polarization of the injected electron spins\cite{doi:10.1063/5.0008988}. If the spins are aligned mainly antiparallel to the intrinsic spin polarization of the ferromagnet, a sufficient spin-orbit torque (SOT) is created and the magnetization is rotated out of its equilibrium direction which is defined by the internal magnetic field in \ce{Ni81F19}(Fig. 1(d)). 
The magnetization starts to precess in the microwave frequency range. 
The Gilbert damping torque (GT) for this precession is compensated by the SOT, resulting in auto-oscillations.
To achieve this compensation, a high current density must be reached which is the underlying reason for forming a constriction in the device (Fig. 1(b)).
Generally, it is believed that large amplitude auto-oscillations are generated at regions with reduced internal magnetic field where the GT is minimized. Auto-oscillations are formed at low frequencies of the dispersion relation in case of in-plane magnetization\cite{ISI:000264630400001} as in our experiment. The smallest frequencies are reached within these magnetic field minima resulting in the formation of spin-wave potential wells (Fig. 1(e)) since the auto-oscillation frequency lies in the spin-wave bandgap of the surrounding material. 

In order to measure the auto-oscillation frequencies in the fabricated samples, all-electrical measurements of the emitted microwave signals are conducted first. These measurements are based on the modulation of the sample resistance via auto-oscillations, affecting the anisotropic magnetoresistance (see Methods and SI 1
). In order to achieve sufficient modulation, the magnetization must be rotated partially in the direction of the dc current. However, this isn't ideal for the generation of auto-oscillations in these structures which prefers the dc current to be perpendicular to the magnetization. Hence, only a small tilting of 23° of the external B field is chosen to ensure the generation of auto- oscillations while simultaneously maintaining a sufficient modulation of the resistance. Since the Gilbert damping has to be compensated by a sufficient spin-orbit torque, a characteristic threshold current has to be exceeded before the auto-oscillation state is achieved. As shown in Fig. 1(f), this current is reached at about $I_{\textnormal{thr}}=6.5$ mA for an externally applied magnetic field of 22.75 mT resulting in oscillations at about 3.4 GHz. 
We observe two oscillation frequencies which decrease with increasing $I_{\textnormal{DC}}$. We attribute this negative frequency shift to several possible effects such as, a) the increased heating of the structure reducing the saturation magnetization, b) the Oersted field that is lowering the effective magnetic field and c) the change of the in-plane magnetization projection during the precession\cite{ISI:000264630400001}. This dc current dependence of the auto-oscillation frequency is an advantageous effect which we use to bring auto-oscillations in resonance to the nitrogen-vacancy (NV) sensor’s spin transition. Note that, no auto-oscillations are observed with reversed polarity of $I_{\textnormal{DC}}$. Due to the symmetry of the spin Hall effect, a reversal of the dc current polarity results in switching of the spin current polarization and the spin-orbit torque. As a result, the damping in the ferromagnet is increased and no auto-oscillations can be excited (see SI 1
).

\subsection*{Mapping magnetic field distribution}
\begin{figure}[t]
\begin{center}
\scalebox{1}{\includegraphics[width=18 cm, clip]{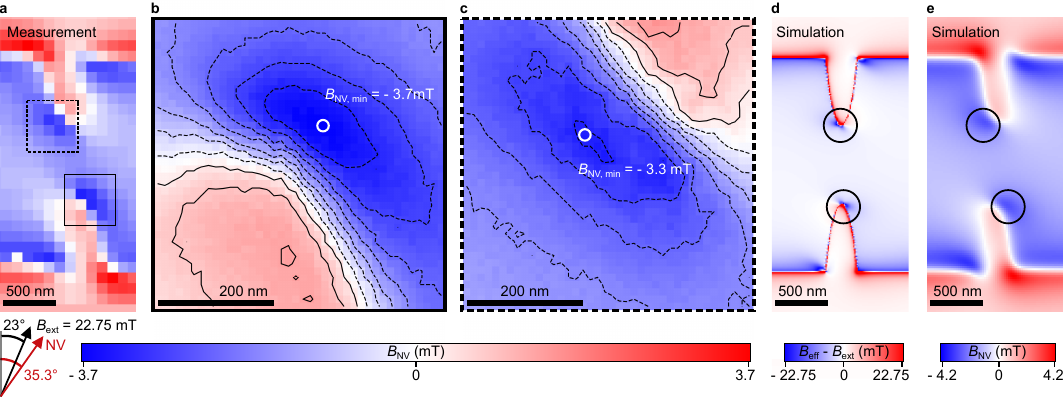}}
\caption{\label{fig2} Magnetic field distribution of the SHNO without dc current (a) Measurement of the magnetic stray field component parallel to the NV axis depicting localized field minima. (b) and (c) High-resolution maps revealing characteristic magnetic field magnitudes at each constriction edge. (d) Simulation of the internal field distribution revealing local field minima at the edges of the constriction (circles). (e) Simulated magnetic stray field component along the NV axis at a height of 80 nm above the SHNO revealing localized field minima (circles) as fingerprint of the minima in (d).}
\end{center}
\end{figure}
NV magnetometry is based on the atomic defect in diamond where two carbon atoms are replaced by a nitrogen (N) atom and a vacancy (V) forming a color center. The negatively charged NV center has a triplet ground state which is characterized by spin dependent fluorescence intensity. In particular, the $m_{\textnormal{S}}=0$ state is characterized by significantly higher fluorescence intensity compared to the $m_{\textnormal{S}}=\pm1$ states. The fluorescence readout can be easily modulated when the transition between these $m_{\textnormal{S}}=0$ and $m_{\textnormal{S}}=\pm1$ states is driven via microwave signal of appropriate frequency\cite{doi:10.1126/science.276.5321.2012}. 

First, we use the sensor in this mode and scan over the switched off SHNO in order to map the generated stray field which provides an insight into the intrinsic magnetic field distribution. We use an AFM tip consisting of a diamond nanopillar with a single NV center oriented in-plane as shown in Fig. 1(a). This ensures that the external magnetic field is mainly applied along the NV spin with a minor misalignment of about 12.3° without affecting the NV resonances and the fluorescence contrast significantly. At each pixel the microwave is applied to a stripline and its frequency is swept to determine the resonance frequency of the NV center to calculate the magnetic stray field. The magnetic field component parallel to the NV axis is shown in Fig. 2(a). We find areas with reduced magnetic field at the constriction edges. Furthermore, the scans with higher spatial resolution shown in Fig. 2(b) and (c) reveal quantitatively different magnitudes of both field minima. 
In order to understand the measurements micromagnetic simulations are conducted to determine the alignment of the magnetic moments inside the SHNO and the effective magnetic field (Fig. 2(d)). We find two localized magnetic field minima at the constriction edges which are results of the strong demagnetization field inside the SHNO constriction. As seen in the measurement of the stray field, the mirror symmetry is broken. This is explained by the orientation of the external magnetic field which controls the position of these internal field minima at the constriction edge. 
The simulated magnetization state is used to calculate the stray field at different heights above the sample. We estimate the distance between NV sensor and sample to about 80 nm by comparing the calculated (shown in Fig. 2(e), see also SI 2
) with the measured stray field distribution and magnitudes in Fig. 2(a). We conclude from this calculation that the minima of the stray field are a fingerprint of the local minima of the effective magnetic field inside the SHNO.
Current knowledge based on only micromagnetic simulations\cite{PhysRevApplied.9.014017,Mutual_synchronization_of_nanoconstriction-based_spin_Hall_nano-oscillators_through_evanescent_and_propagating_spin_waves,doi:10.1063/1.4901027} suggests that auto-oscillations can be trapped in these minima that act as spin-wave potential wells and, therefore, the different field magnitudes explain the generation of auto-oscillation modes with distinct frequencies.

\subsection*{NV-spin manipulation via SHNO microwave field}
Recently, the interaction of the microwave field descending from spin waves with spin defects in diamond or silicon carbide was demonstrated\cite{Iacopo:uz, doi:10.1126/science.adj7576, Simon:2022aa, Simon:2021aa, PhysRevB.102.024404, Lee-Wong:2020vc, Sar:2015tr, doi:10.1126/sciadv.abg8562, Yan_2022, Wang:2020vf, doi:10.1126/sciadv.adi2042,doi:10.1073/pnas.2019473118,Andrich:2017aa,PhysRevB.89.180406}. 
In order to verify the quality of the auto-oscillation modes as miniaturized MW sources, we now drive the NV sensor on the scanning tip in resonance by only activating the SHNO device. Thereby, no external microwave power but a dc current is applied to the SHNO to generate magnetic auto-oscillations. Due to the precession of the magnetic moments, the dipolar magnetic field is modulated with the frequency of the auto-oscillations in the gigahertz range in close vicinity to the sample. By scanning the diamond AFM tip over the auto-oscillation, the fluorescence intensity of the NV center drops as a function of the generated microwave power only when the SHNO is tuned to provide the desired NV resonance frequency. Therefore, the distribution of the magnetic auto-oscillations can be acquired by simply measuring the fluorescence intensity of the NV sensor. Contrary to a standard spin resonance measurement where excitation frequency or external field sweeps are required, our method is significantly faster as it only relies on our photon collection efficiency (100 ms/pixel). 

\begin{figure}[t]
\begin{center}
\scalebox{1}{\includegraphics[width=8.8 cm, clip]{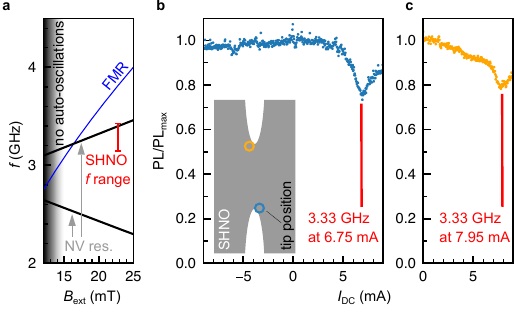}}
\caption{\label{fig3} NV-spin manipulation via SHNO microwave field. (a) The calculated NV resonance (black) at higher frequency overlaps with the SHNO frequency range (red) extracted from Fig. 1(f). This auto-oscillation range is located below the calculated thin-film FMR range (blue) as expected for SHNOs in this measurement geometry. 
(b) NV photoluminescence (PL) drops by 24 \% when the SHNO is activated by a characteristic positive DC current of $I_{\textnormal{DC}} = 6.75$ mA. The PL reduces only for positive currents for which the SHNO is active. The lowest PL is reached when the SHNO microwave oscillations are in resonance with the NV at 3.33 GHz. The inset shows the position of the NV-AFM tip during the dc sweep.
(c) NV photoluminescence (PL) drops by 21 \% at the second constriction edge when the SHNO is activated by a characteristic positive dc current of $I_{\textnormal{DC}} = 7.95$ mA. }
\end{center}
\end{figure}
Since the frequencies of the NV transition and the auto-oscillations depend on the external magnetic field, there is a limited parameter space allowing synchronization of both. Note that, the SHNO auto-oscillation frequency can be controlled by the applied dc current. Whereas it only slightly influences the NV sensor due to the change of the Oersted field above the sample. Figure 3(a) depicts the calculated NV resonance frequencies in black, the calculated ferromagnetic resonance (FMR) frequency of a 5 nm thick \ce{Ni81F19} thin film in blue and the measured frequency range of the SHNO device in red. At magnetic fields below $B_{\textnormal{ext}} = 15$ mT no auto-oscillations are expected due to incomplete alignment of the magnetic moments in the SHNO caused by the shape anisotropy. Due to the demagnetization field within the constriction (Fig. 2) and the nonlinear redshift of the auto-oscillations (Fig. 1(c)), the auto-oscillations are located at frequencies below the FMR. As a result, a crossing of the auto-oscillation frequency range and the upper NV resonance exists at the used magnetic field of $B_{\textnormal{ext}} = 22.75$ mT. Notably, besides this frequency matching condition, it is imperative to have sufficient microwave power generated by the auto-oscillations at the position of the NV sensor in order to have any interaction between the sensor and our SHNO device. 

In order to prove this, the diamond tip containing the NV center is brought in contact with the SHNO at the bottom constriction edge as shown in the inset of Fig. 3(b). 
The dc current is swept stepwise from -9 mA to 9 mA and the emitted fluorescence is acquired simultaneously. 
Close to the estimated NV resonance frequency determined from Fig. 3(a) the fluorescence intensity decreases strongly and forms a characteristic dip. We attribute this to the fingerprint of a specific NV spin transition ($m_{S}=0$ to $m_{S}=+1$), driven by the microwave field generated by the auto-oscillations inside the SHNO. The minimum is reached at $I_{\textnormal{DC}} = 6.75$ mA which corresponds to a microwave frequency of 3.33 GHz determined from Fig. 1(f). 
The acquired fluorescence for the sweep with the reversed dc current polarity doesn't show a characteristic dip, which corroborates with the electrical measurement of the auto-oscillations (SI 1). 
After finding the second resonance condition of $I_{\textnormal{DC}} = 7.95$ mA at the second constriction edge (Fig. 3(c)) these magnitudes are fixed, respectively and lateral scans of the NV tip are conducted in order to acquire the fluorescence output at each pixel. This results in maps showing strong reductions of the fluorescence intensity at positions where the microwave field and, therefore, the auto-oscillation modes are present in the SHNO.

\subsection*{Mapping the auto-oscillation modes}
\begin{figure}[t]
\begin{center}
\scalebox{1}{\includegraphics[width=18 cm, clip]{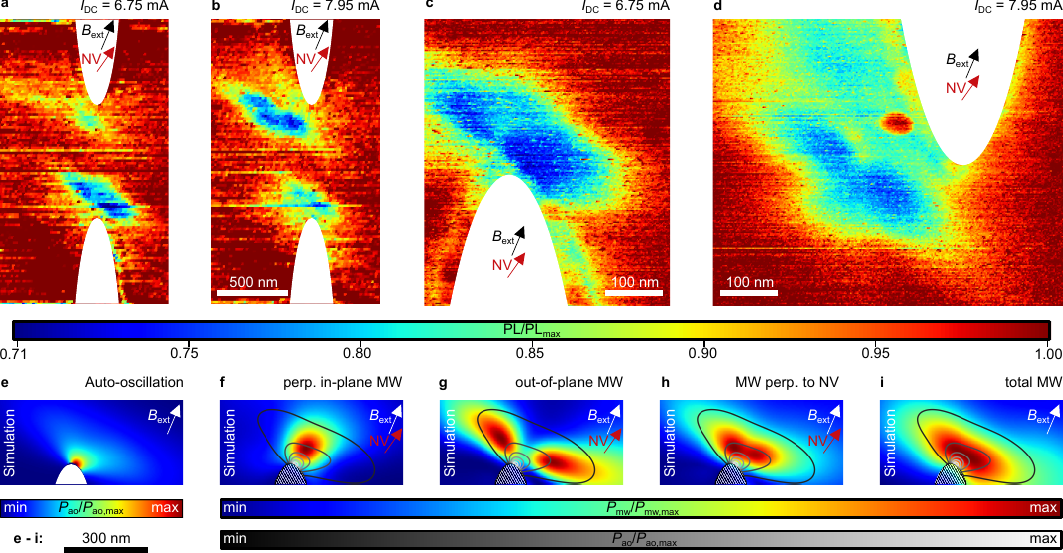}}
\caption{\label{fig4} Localization of the auto-oscillation modes (a), (b) PL map at $I_{\textnormal{DC}} = 6.75$ mA ($I_{\textnormal{DC}} = 7.95$ mA) when auto-oscillation Mode 2 (Mode 1) is in resonance with the NV at the bottom constriction edge (upper constriction edge). For both dc currents only one of the modes is in resonance with the NV and the second one off-resonance, respectively. (c), (d) Finer PL map at the interaction area of Mode 2 (Mode 1) revealing two lobes. (e) Micromagnetic simulation of the auto-oscillation power localized at the constriction edge. (f), (g) Microwave field in-plane (out-of-plane) component being perpendicular to the NV axis. (h) Microwave field interacting with the NV revealing two lobes as seen in the experiment. (i) Total microwave field (Microwave fields are calculated at a height of 80 nm above the SHNO).}
\end{center}
\end{figure}
%
First, a reference map at $I_{\textnormal{DC}} = 0$ mA is obtained which is substracted from the maps at nonzero dc currents. As no auto-oscillation are present in this condition, it serves as a background to eliminate any other spurious effects that may also influence the photon count rate. At $I_{\textnormal{DC}} = 0$  mA, only a slight variation in the PL is recorded (compare SI 3
), which changes dramatically for the PL maps taken at $I_{\textnormal{DC}} = 6.75$ mA and $I_{\textnormal{DC}} = 7.95$ mA (Fig. 4(a), (b)). For 6.75 mA an additional feature appears at the bottom constriction edge along with a moderate variation at the top edge. For 7.95 mA however, a strong feature appears at the top edge and the one at the bottom edge almost vanishes. Figures 4(c) and (d) capture these areas in more detail. This behavior is attributed to the two auto-oscillation modes present in this sample. As shown in Fig. 1(c), these modes have distinct frequencies and, therefore, different dc currents have to be applied to the SHNO to tune their frequency to the NV transition. Since the second mode has a lower frequency than the first one, a smaller dc current has to be applied to drive the NV resonance. Hence, Mode 2 is located at the bottom edge of the constriction and Mode 1 is located at the top edge. This conclusion is supported by the quantitative measurement of the stray field shown in Fig. 2(b) and (c). A smaller stray field is measured at the bottom edge of the constriction caused by the local internal magnetic field minimum which results in a lower auto-oscillation frequency as determined for Mode 2. Both internal field minima form spin-wave potential wells (see Fig. 1(e)). Due to the spin-wave band gap at low frequencies, the auto-oscillations can't excite propagating waves in the surrounding area. Therefore, the area of reduced magnetic field at the constriction edges acts as a resonator where high auto-oscillation amplitudes are reached. Additionally, the locally reduced fields result in a smaller Gilbert damping torque which is another contribution for increased precession angles. Due to the formation of two local field minima, a high power and coherent single-mode microwave signal can only be reached if both auto-oscillations can synchronize mutually. This is possible in case of a sufficiently small frequency difference. 

Interestingly, the PL maps shown in Fig. 4(c) and (d) reveal two lobes within the area of NV-SHNO interaction which is a result of the generated microwave field of the auto-oscillations. In order to understand this shape, micromagnetic simulations are conducted. As shown in Fig. 4(e), the strongest auto-oscillation intensity is formed at the constriction edge which is in agreement with our experimental observation. This data was used to calculate the microwave power at a height of 80 nm above the SHNO. Fig. 4(f) and (g) show the in-plane and the out-of-plane component of this microwave field respectively, which are both perpendicular to the NV axis. Surprisingly, the strongest microwave field isn't generated above the maximum of the auto-oscillation power. Additionally, the out-of-plane component has two microwave power maxima. We attribute these behaviors to the elliptical precession cone of the magnetic moments in this thin film sample (see SI 4
). Fig. 4(h) shows the combination of the perpendicular components to the microwave field which interacts with the NV. There, two maxima are visible which form lobes similar to the feature seen in the measurement. The simulation allows us to conclude that the spot of highest auto-oscillation intensity is located between these two lobes and the constriction edge. For comparison, the total microwave power is shown in Fig. 4.(i), which doesn't show the two lobes. Hence, the lobes are seen due to the selective interaction of the NV with the three-dimensional microwave field.
\newline 
To the best of our knowledge, this is the first nanoscale imaging of the magnetization dynamics in a ferromagnetic metal utilizing a near-surface single NV-center. Furthermore, we demonstrate that such nano-oscillator devices produce a strongly localized microwave field, which might be attractive as a miniaturized microwave source for specific quantum technologies where a propagating or global microwave field may not be preferred. The combination of nanoscale resolution and quantitative magnetometry demonstrated in this work, uniquely determines the magnetic field distribution of a real auto-oscillator device. Quantum sensing of the microwave field generated by the auto-oscillations revealed their position at two separated local minima of the magnetic field. This confinement is explained by the formation of spin wave potential wells due to the local lowering of the spin-wave band gap which prohibits propagating spin waves. The quantitative measurement revealed the distinct stray magnetic field magnitudes at these potential wells explaining the formation of two auto-oscillations modes with specific frequencies. The asymmetry of the auto-oscillation spots with respect to the sample symmetry is caused by the position of these potential wells controlled by the external magnetic field(SI 5
). This proves the importance of local field minima for the formation of auto-oscillations instead of being defined by the areas of largest antidamping only. This work opens up new possibilities with more profound insights about how to control multimode auto-oscillations in real devices where the coherence and output power are currently limited. Our work and further investigations to reveal dynamical properties of such devices will pave the way to design new SHNO or auto-oscillator devices with engineered internal magnetic field distributions.


%
%

%
%

\subsection*{Methods}
\textbf{SHNO fabrication:} The shape of the SHNO and the electrical contacts were fabricated using electron-beam lithography. The metallic layers of the SHNO were deposited using magnetron-sputtering. The bottom Ta layer is used as seed layer and the top one for oxidation protection. The gold for the electrical contacts was deposited using thermal evaporation. A 5 nm thick Cr layer was used as adhesion layer below the 100 nm thick Au layer. 

\textbf{Electrical measurements:} A dc current source was used to supply the direct current. It was passed through a bias tee, a microwave probe and impedance-matched electrical contacts to the SHNO. The auto-oscillations generate a microwave signal $I_{\textnormal{MW}}$ via the anisotropic magneto-resistance within the \ce{Ni81F19} layer. This signal is transmitted through the microwave probe and passes the bias tee through the high frequency output. Three low-noise microwave-amplifiers amplify the signal by about 65 dB before the signal is acquired by a spectrum analyzer. 

\textbf{NV quantum sensing:} We used a commercially available diamond tip (Qzabre, Q5) with a single NV in in-plane orientation (110 cut) in order to minimize components of the external magnetic field being perpendicular to the NV axis. We use this sensor in two different measurement modes. 1) Spatially-resolved NV spin resonance: In order to determine stray magnetic field at each pixel above the sample quantitatively, the SHNO was switched off and the microwave was applied to an separate stripline at various frequencies. At the resonance frequency the fluorescence intensity of the NV dropped and this frequency was used to calculate the magnetic field component parallel to the NV axis. 2) Spatially-resolved auto-oscillation detection: The external microwave is switched off. The microwave field is generated by the auto-oscillations within the SHNO which causes the NV spin transition. During the lateral scan, the NV fluorescence would drop only at pixels that are in close vicinity to the auto-oscillations. Therefore, the locations of auto-oscillations are revealed relatively fast compared to other established methods because only the fluorescence intensity has to be acquired once at each pixel (100 ms/pixel). 

\textbf{Micromagnetic simulations:} First, the SHNO was modeled in COMSOL Multiphysics. The dc current density in the Pt layer and the resulting Oersted field within the \ce{Ni81F19} layer were simulated using the conductivities $\sigma_{\textnormal{Pt}}=3.1$ MS/m and $\sigma_{\textnormal{\ce{Ni81F19}}}=1$ MS/m. 
The shape, dc current density and Oersted field were exported and used as input for the micromagnetic simulations. Mumax3\cite{MuMax3} was used to study the relaxed magnetization state and the time-evolution of the dynamic magnetization in the SHNO. An external field of 22.75 mT is applied under 23° as shown in Fig. 1(a). The saturation magnetization, exchange stiffness and Gilbert damping parameter were set to $M_{\textnormal{sat}}=630$ kA/m, $A_{\textnormal{ex}}=10$ pJ/m and $\alpha =0.02$. A polarization factor of $P=0.16$ was used to convert the imported current density to the spin-current density. The orientation of the spin-current polarization is perpendicular to the current density in each pixel of the simulation.  A simulation area of 2000 nm x 2000 nm x 5 nm around the constriction was discretized in 512 x 512 x 1 points. A time-window of 200 ns was simulated with 50 ps time steps. The FFT revealed the auto-oscillation frequency. The spatial distribution of this auto-oscillation intensity was plotted in Fig. 4(e). In order to achieve two distinct auto-oscillation frequencies as in the experiment, a small asymmetry was introduced between both constriction edges (see SI 6
).

\textbf{Microwave field calculations:} The spatial distribution of the dynamic magnetization from the micromagnetic simulations was used to calculate the stray magnetic field at 80 nm (distance to NV sensor) above the SHNO for each time step. The FFT revealed the dynamic three-dimensional microwave magnetic field at the NV position caused by the auto-oscillations. Fig. 4(f)-(i) show the squared magnitudes being proportional to the generated microwave power.  

\subsection*{Acknowledgments}
A.S. gratefully acknowledges an IQST-YR grant from the Center for Integrated Quantum Science and Technology supported by funding from the Carl Zeiss Foundation and an Emmy Noether grant from the Deutsche Forschungsgemeinschaft (project no. 504973613). J.W. acknowledges: BMBF via project QCOMP, the EU via project AMADEUS and the DFG via GRK2642. We gratefully acknowledge Frank Thiele and Gunther Richter from the Central Scientific Facility Materials at the Max Planck Institut for Intelligent Systems (Stuttgart) for the thin film deposition. 

\subsection*{Author contribution}
T.H. and A.S. envisioned and designed the experiment. T.H. and A.A. built the electrical measurement setup. T.H. 

fabricated the SHNO devices. T.H. and A.A. performed electrical characterization of the SHNO devices. T.H. conducted the NV measurements, performed the micromagnetic simulations as well as microwave field calculations. T.H. and A.S. prepared the manuscript with contributions from all coauthors.

\bibliography{Paper-data-base}

\end{document}


\title[Article Title]{Supplementary Information for 

Nanoscale Mapping of Magnetic Auto-oscillations with a single Spin Sensor}


\author*[1,2]{\fnm{Toni} \sur{Hache}}\email{t.hache@fkf.mpg.de}

\author[1]{\fnm{Anshu} \sur{Anshu}}

\author[2]{\fnm{Tetyana} \sur{Shalomayeva}}



\author[2]{\fnm{Rainer} \sur{Stöhr}}

\author[1,3]{\fnm{Klaus} \sur{Kern}}

\author[1,2,4]{\fnm{Jörg} \sur{Wrachtrup}}

\author*[1,4]{\fnm{Aparajita} \sur{Singha}}\email{a.singha@fkf.mpg.de}

\affil*[1]{\orgname{Max Planck Institute for Solid State Research}, \orgaddress{\street{Heisenbergstr. 1}, \city{Stuttgart}, \postcode{70569}, \country{Germany}}}

\affil[2]{\orgdiv{3rd Institute of Physics and Research Center SCoPE}, \orgname{University of Stuttgart},  \orgaddress{\city{Stuttgart}, \postcode{70049}, \country{Germany}}}

\affil[3]{\orgdiv{Institute de Physique}, \orgname{École Polytechnique Fédérale de Lausanne}, \orgaddress{\city{Lausanne}, \postcode{CH-1015}, \country{Switzerland}}}

\affil[4]{\orgname{Center for Integrated Quantum Science and Technology IQST, University of Stuttgart, \city{Stuttgart}, \postcode{70049}, \country{Germany}}}


\maketitle

\section{Electrical measurements of Auto-oscillations} 
In order to extract the auto-oscillation signals from the all-electrical measurements (Fig. 1(f) of the main manuscript) a background subtraction algorithm was used. After acquiring the data by the spectrum analyzer as function of the dc current, each frequency channel was fitted to a quadratic function with subsequent subtraction (Fig. SI 1(a)). This removes signals stemming from standing waves within the microwave circuit and the noise resulting from Joule heating. Figure SI 1(b) shows that auto-oscillation are excitable only at positive dc currents due to the symmetry of the spin Hall effect. A reversal of the dc current polarity results in the switching of the spin current polarity. This flips the the direction of the spin-orbit torque which results to an increased damping in the system and, therefore, no auto-oscillations can be generated. Figure SI 1(c) shows a single spectrum of SI 1(b) at 8.5 mA.
\begin{figure}[h]
\begin{center}
\scalebox{1}{\includegraphics[width=18cm, clip]{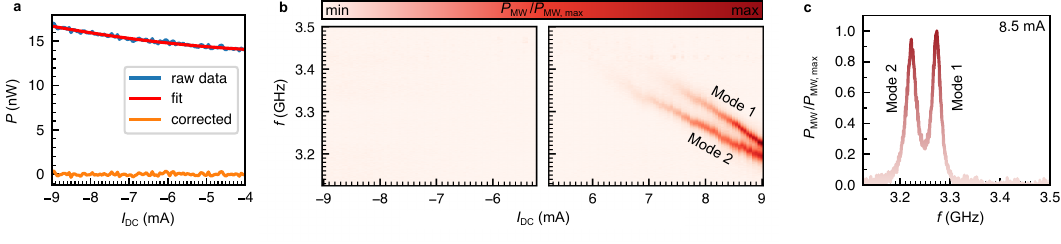}}
\captionsetup{labelformat=empty}
\caption{\newline\textbf{Figure SI 1:} a) Example of background subtraction from the all-electrical measurement. The increase of the background noise level as a function of dc current is a result of the increasing Joule heating. b) Generation of auto-oscillations is achieved only for positive dc currents. At negative currents, the sign of the SOT is reversed and the damping in the system is increased prohibiting the formation of auto-oscillations. c) Single spectrum at 8.5 mA.}
\end{center}
\end{figure}

\section{Height dependence of the magnetic stray field component parallel to the NV axis} 
We estimated the distance of the NV sensor to the SHNO by comparing shape and values of the calculated magnetic stray field to the measured distribution shown in Fig. 2 of the main manuscript. Figure SI 2 shows the calculated magnetic stray field generated by the \ce{Ni81F19} inside the SHNO at different heights. 
\begin{figure}[h]
\begin{center}
\scalebox{1}{\includegraphics[width=18cm, clip]{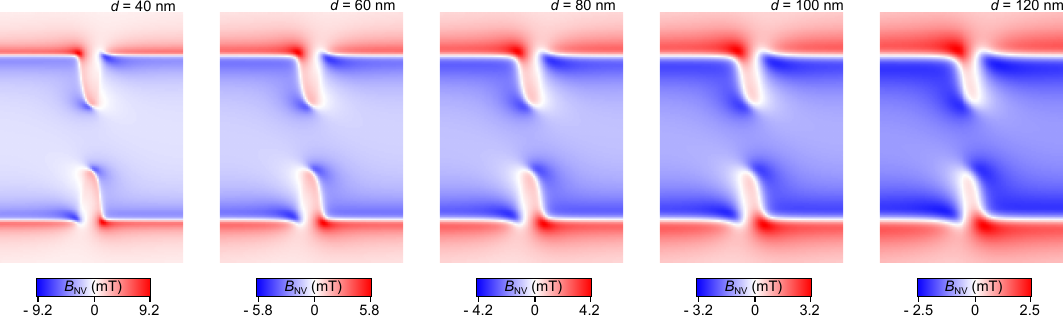}}
\captionsetup{labelformat=empty}
\caption{\newline\textbf{Figure SI 2:} Calculated magnetic stray field along the NV axis at different distances above the SHNO.}
\end{center}
\end{figure}
The contrast between magnetic field minima within the constriction and the surrounding area becomes smaller with increasing distance from the sample. The shape and the values of the magnetic field at 80 nm fits best to the quantified stray field in the experiment. 

\section{PL maps before background subtraction} 
The measurement data shown in Fig. 4 of the main manuscript was obtained after two evaluation steps. 1) Normalization of the photon count rate of each row by an averaged value outside of the constriction area in order to exhibit the PL contrast on the maps. 2) Subtraction of the PL map without applied dc current to remove any spurious other effect influencing the photon count rate apart from the auto-oscillations. 
Figure SI 3(a) shows the mapped photon count rate at each pixel at $I_{\textnormal{DC}}= 0$ mA. There is a slight reduction of the photon count rate at the top left area which isn't caused by magnetic auto-oscillations and, therefore, this map is subtracted from the maps with auto-oscillations. Figure SI 3(b) and (c) show the PL maps (before background subtraction) when the auto-oscillations are active.  
\begin{figure}[h]
\begin{center}
\scalebox{1}{\includegraphics[width=8.8cm, clip]{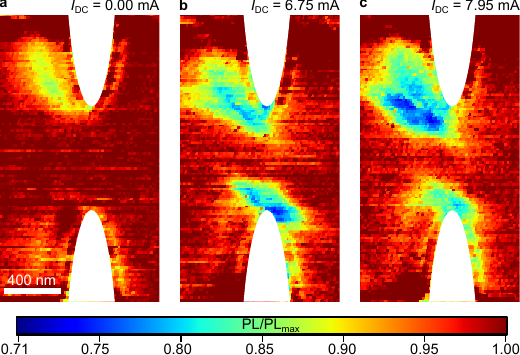}}
\captionsetup{labelformat=empty}
\caption{\newline\textbf{Figure SI 3:} Measured PL maps before background substraction. (a) Background map at $I_{\textnormal{DC}}= 0$ mA with a slightly reduced photon count rate at the top left area. (b), (c) PL maps (before background subtraction) when the auto-oscillations are active.}
\end{center}
\end{figure}
%

\section{Microwave field components generated by a single precessing spin} 
In order to get a deeper understanding of the microwave field generation caused by the auto-oscillations (Fig. 4 f) to i)) we calculated the microwave generation 80 nm above a single precessing spin. For each time step, the dipolar magnetic field was calculated resulting in time-dependent dipolar field data. The FFT of this time-dependent field gives the dynamic components at the precession frequency. 
\begin{figure}[h]
\begin{center}
\scalebox{1}{\includegraphics[width=18cm, clip]{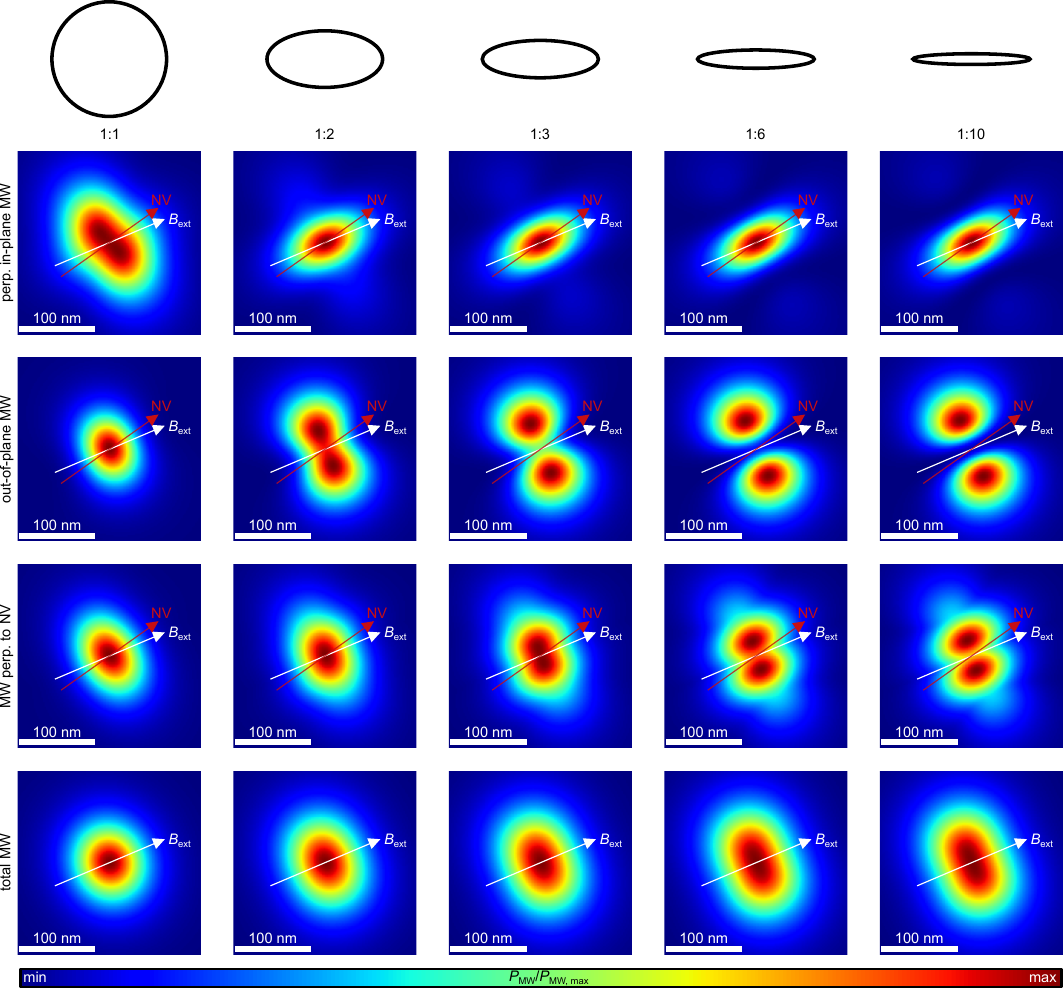}}
\captionsetup{labelformat=empty}
\caption{\newline\textbf{Figure SI 4:} Calculated microwave field components of a single precessing magnetic moment calculated at a distance of 80 nm. Arrows indicate the alignment of the magnetic field and the NV axis which correspond to the angles in the experiment. }
\end{center}
\end{figure}
To emulate the elliptical precession of the magnetic moment in a thin film sample caused by the shape anisotropy, different ratios of the minor and major axis of the ellipse from 1:1 to 1:10 are used (first row of Fig. SI 4). With increasing ellipticity of the precession the out-of-plane component of the microwave field shows an increasing separation of two maxima. The in-plane component shows always only one maximum independent of the ellipticity of the precession cone. The total microwave field which can interact with the NV shows similar to the out-of-plane component two maxima with increasing separation for stronger ellipticity. The total microwave field power has always only one maximum. The shape of the total microwave power distribution becomes more anisotropic for increased ellipticity. The elliptical precession with a ratio 1:6 correspond to the precession behavior of the auto-oscillating magnetic moments shown in Fig. 4(e) in the main manuscript. Based on this calculations we conclude that the formation of two maxima in Fig. 4 g) and h) of the main manuscript is caused by the elliptical precession of the magnetic moments. 
\section{Control of auto-oscillation areas via external magnetic field angle} 
Micromagnetic simulations are conducted to investigate the dependence of the auto-oscillation formation on the position of the internal magnetic field minima (spin-wave potential wells) in \ce{Ni81F19}. Figure SI 5(a) shows how the magnetic field minima shift along the edge of the constriction to the right when the the external magnetic field angle is increased from 0° to 60°. Figure SI 5(b) shows the position of the auto-oscillations at the constriction edge at these field angles, respectively. It can be clearly seen, that the auto-oscillation spots follow the internal magnetic field minima. As mentioned in the main manuscript, the asymmetry of the measured auto-oscillation spots with respect to the sample geometry is caused by the shift of the internal magnetic field minima defined by the external magnetic field angle.  
\begin{figure}[h]
\begin{center}
\scalebox{1}{\includegraphics[width=18cm, clip]{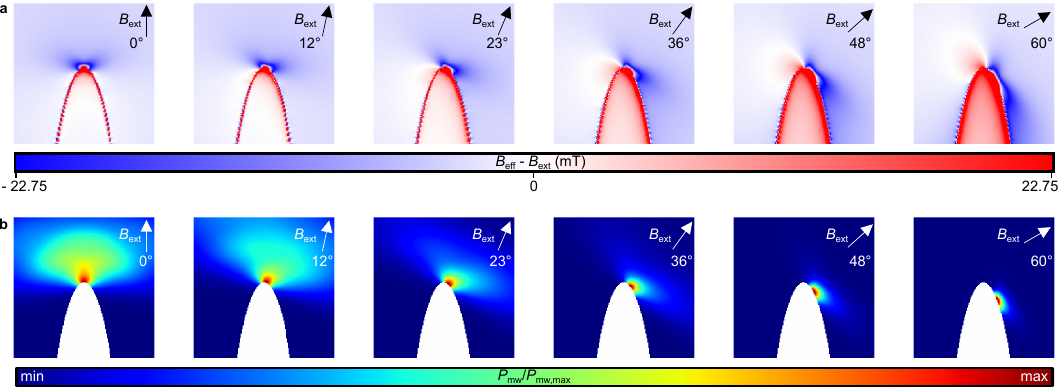}}
\captionsetup{labelformat=empty}
\caption{\newline\textbf{Figure SI 5:} The in-plane angle of the external magnetic field defines the position of the internal magnetic field minima as shown in (a). The auto-oscillation areas shown in (b) follow the position of these internal magnetic field minima which act as potential wells for spin-waves.}
\end{center}
\end{figure}
\newpage
\section{Shape of the SHNO for the micromagnetic simulation} 
A micromagnetic simulation is conducted in order to analyze the microwave field generated above. A small asymmetry is introduced in order to emulate the behavior in the experiment for which both auto-oscillation modes at the edges show different frequencies and, therefore aren't mutually synchronized. The SHNO shape is shown in Fig. SI 6. The asymmetry leads to slightly different auto-oscillation frequencies of 3.62 GHz at the bottom constriction edge and 3.65 GHz at the top constriction edge. The auto-oscillation mode profile at the bottom constriction is shown in Fig. SI 6. This mode is shown in Fig. 4(e) of the main manuscript and was used as input for the microwave field calculations shown in Fig. 4(f) to (i). This simulation reveals that already small deviations from the mirror symmetry of the constriction has a strong influence on the overall spectral characteristic of the device. In the experiment, this might be amplified by variations in the saturation magnetization, for instance, due to different oxidation at the edges.

\begin{figure}[h]
\begin{center}
\scalebox{1}{\includegraphics[width=8.8cm, clip]{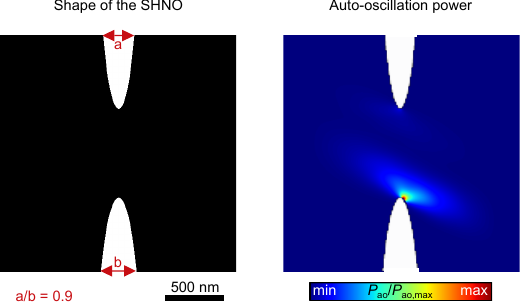}}
\captionsetup{labelformat=empty}
\caption{\newline\textbf{Figure SI 6:} The asymmetric shape leads to distinct auto-oscillation frequencies at both constriction edges. The auto-oscillations at 3.62 GHz are located at the bottom constriction edge with the shown mode profile.}
\end{center}
\end{figure}
%

\bibliography{Paper-data-base}